\documentclass[11pt,]{amsart}   	
\usepackage{geometry}                		
\usepackage{graphicx}				
\usepackage{amssymb}
\usepackage{epsf}
\usepackage{amsmath}
\usepackage[running]{lineno}
\title{Fitting theory to data in the presence of
background uncertainties}
\author{Byron Roe}
\address{University of Michigan, Ann Arbor, MI 48105}

\begin{document}

%
\mbox{}\\
{\bf BooNE Technical Note 304}
\vskip 20mm
%
\today

%
%
%
\begin{abstract}
When fitting theory to data in the 
presence of background uncertainties, the question of whether the spectral shape
of the background happens to be similar to that of the theoretical model 
of physical interest has not generally
been considered previously.  These correlations in shape are considered in the
present note and found to make important corrections to the calculations.
The discussion is phrased in terms of $\chi^2$ fits, but the general
considerations apply to any fits.
Including these
new correlations provides a more powerful test for confidence regions. 
Fake data studies, as used at present, 
may not be optimum.  
Neutrino oscillations are used as examples, but the problems
discussed here are general ones.
\end{abstract}

\maketitle
\section{Introduction}
When fitting theory to data in the 
presence of background uncertainties, the question of whether the spectral shape
of the background is similar to that of the theoretical model 
of physical interest has not generally
been considered previously.  These correlations in shape are considered in the
present note and found to make important corrections to the calculations.

Some causal correlations between background and the theoretical model are
usually included at present.
For example, beam normalization uncertainties affect
both the theoretical model and the background. 
There are also some correlations when
the size of the theoretical model affects the size of the background.

However, when the theoretical model parameters are allowed to vary in a fit,
there is a new, qualitatively different, correlation 
that should be considered if the error matrix includes the effects of
uncertainties in the backgrounds.  
  
This correlation is not a causal correlation, but occurs simply
due to a similarity between 
the spectrum of the theoretical model over the
bins and the spectrum of the backgrounds.
This similarity can be quantified using a
regression analysis.  These correlations should, perhaps, be
distinguished by being called ``regression correlations''. 

Including these
new correlations provides a more powerful test for confidence regions. 
Use of Monte Carlo simulations to estimate confidence regions (``fake data
studies'') as described in Section 6 give valid confidence regions
with or without the corrections described here.  However, the use
of these correlations allows a more powerful test, often giving smaller
regions for a given confidence level.  This effect will be illustrated in the
numerical example studied in Section 7.

A chisquare fit will often not have chisquare distribution with or without the
corrections here. The problem, which is well, but sometimes not widely
known, is that systematic background errors are really different than
the statistical errors.  If an experiment is repeated, the numbers of
events are different each time because of the statistical errors.
However, the background expectation is the same each time.  Nature has
chosen some value.  The systematic error here is the uncertainty in
our knowledge of what is that value, a Bayesian concept which does
not fit comfortably in the frequentist $\chi^2$ method.  
When this systematic effect is included in the
error matrix as if it were a statistical error, 
a chisquare probability may not result.  This point will also be
illustrated in the numerical example in Section 7.

The toy
model considered in Section 3 describes the essence of the problem
considered in this note.

\section{Notation}

Suppose data have been obtained for a histogram with $N_{bins}$.  
The $i$th bin has $N_{data}(i)$ events.
The model used for fitting the data has background parameters and theory
parameters.  
 
Let $B_j(i)$ be the true expected value of background $j$ in bin $i$.  
Let $N_{bkrd}$ be the number of background
parameters associated with these backgrounds.  For any
given data set, the number of background events will have a statistical
spread with mean value $B_j(i)$.  
 It is assumed that the backgrounds have
been evaluated in independent 
experiments previously, giving  estimates for the
mean value of each $(B_m)_j(i)$ and the covariance matrix of the error in the
estimated mean value $cov_{(B_m)_j}$.  
There are correlations from bin to bin 
and also correlations between the various backgrounds, 
so that the
covariance matrix is not, in general, diagonal. 
The total estimated background expectation is $(B_m)_{tot}(i)$.

For the initial discussion, it is assumed that there are
a string of separate backgrounds, each with one parameter; the total
background is then    
\begin{linenomath*}
\begin{align}
(B_m)_{tot}(i) =  \sum_{j=1}^{N_{bkrd}} (B_m)_j(i).
\end{align}
\end{linenomath*}
Each background has the form
$B_j(i)=b_jg_j(i)$, where
$i= 1,\cdots,N_{bins}$, and $j= 1,\cdots,N_{bkrd}$.
The $b_j$
correspond to scale parameters defining the level of background.
The bin dependence
of each of the backgrounds $g_j(i)$ is assumed known and fixed.
In Section 5, the form of the backgrounds will be generalized.

The second kind of parameters are parameters of the theoretical model of
physical interest and are
initially unknown. 
There
are $N_t$ theory parameters, $t_k,\ k=1,\cdots,N_t$.  
The theory model prediction
for each bin, $N_{th}(i)$, is a function of these parameters.

Define the ``signal'' as the data minus the background estimates.
This is the part of the data that will be fitted to the theoretical model
of interest.
\begin{linenomath*}
\begin{align} 
N_{sig}(i) \equiv N_{data}(i) - (B_m)_{tot}(i).
\end{align}
\end{linenomath*}

Two common $\chi^2$ fitting methods are considered.
For the first fitting 
method, the
parameters assigned to the background(s) are held fixed in the fit at their
measured values. 
The uncertainty in
the background parameters enters into the covariance used in the fit.
The covariance matrix of the signal in bins $i,j$ is given as 
\begin{linenomath*}
\begin{align}
cov_{sig}(i,j) = cov_{data}(i,j) + cov_{(B_m)_{tot}}(i,j).
\end{align}
\end{linenomath*}

$cov_{data}(i,j)$ is diagonal and is the variance expected for the data 
in each bin given the theory plus background prediction for the mean number of
events in that bin.

The fit done is to minimize the $\chi^2$, 
\begin{linenomath*}
\begin{align}
\chi^2 = \sum_{i,j=1}^{N_{bins}} [N_{sig}(i)-N_{th}(i)][cov^{-1}_{sig}(i,j)]
[N_{sig}(j)-N_{th}(j)],
\end{align}
\end{linenomath*}
where $[cov^{-1}_{sig}(i,j)]$ is the inverse of the signal covariance matrix.

A second approach uses a constrained fit to account for
backgrounds, 
constraining the backgrounds to be at their measured values
within errors, when performing fits.  The $\chi^2$
is written as a $\chi^2$ ignoring the background uncertainties and
a sum of terms (``pulls'') for each background.  For simplicity, 
assume for the initial discussion 
that there are no correlations between backgrounds and that
each background depends on a single variable. 
Let $\beta_{k}(i)
= \frac{\partial (B_{tot}(i)+N_{th}(i))}{\partial\delta_k}|_{\delta_k=0}$, 
where $\delta_k$ is
the deviation of the true $k$th background or nuisance parameter 
from its measured value. The set of $b_k$ parameters
is usually enlarged for this fit method 
to include nuisance parameters which can affect both the theory
and the backgrounds, such as the overall normalization of the data.
However, in this note the $\delta_k$ will refer only to the background 
parameters.
The experimental quantities are $N_{data}(i)$ and $(B_m)_k(i)$, 
the latter obtained from
previous experiments. $\beta_{k}(i)$ is a
fixed quantity obtained from the spectrum of the backgrounds.
(For the initial assumption that $g_k(i)$ is known, $\beta_k(i)=g_k(i)$.)

$N_{th}$ and the $\delta_k$ are quantities to be fitted.
Let $\sigma^2_{b_k}$ be the variance of the estimate 
of background parameter $b_k$.
\begin{linenomath*}
\begin{align}
\chi^2 = \sum_{i=1}^{N_{bins}}\frac{ \big[N_{data}(i)-\big(N_{th}(i)+\sum_k((B_m)_k(i)
+\beta_{k}(i)\delta_k)\big)\big]^2}{\sigma^2_{data}(i)}
+\sum_k\frac{\delta_{k}^2}{\sigma^2_{b_k}}.
\end{align}
\end{linenomath*}
In the presence of correlations between backgrounds, the usual introduction 
of the covariance matrix is used.
Since each pull term introduces an effective extra bin and an extra
parameter to be fit, the number of degrees of freedom is unaffected.
If one performs Gaussian integration in Equation 5 over the $\delta_k$, 
this approach reduces to the 
previous one~\cite{stump}.

As indicated in the introduction, the uncertainty in the background 
 is the uncertainty 
in our knowledge of that value, a Bayesian concept.  The constrained
fit treats the background expectation value as another parameter
to be fitted along with the theoretical parameters.

$\chi^2$ fits are often done in an iterative fashion taking one step
at a time looking for the  minimum.  The correlations are re-evaluated
at each step.

\section{A toy model}
Suppose one is fitting a theoretical model parameter to data, where the
theoretical model is
of the  form $N_{th}(i)=tf(i)$, where $t$ is a constant to be fit
and $f(i)$ is
a known function of the bin number $i$. Suppose, further, that there is
a single background which is uncorrelated with the
model parameter.
$B(i)=bg(i)$ with $g(i)$ known. Let $b_m$  be the 
estimated mean of $b$ and $\sigma^2_{b_m}$ be the estimated variance of $b_m$. 
The fit is done by using the method of Equations 3 and 4. 
The covariance of the background is $cov_{B_m}(i,j) =g(i)g(j)\sigma^2_{b_m}$. 
The overall covariance, $cov_{overall}$,
is the sum of  the statistical 
variance of the data, $\sigma^2_{data}$ and the covariance of the 
background, $cov_{overall}(i,j)=\sigma^2_{data}(i)\delta_{ij}+cov_{B_m}(i,j)$. 
\begin{linenomath*}
\begin{align}
\chi^2 =
 \sum_{i=1}^{N_{bins}}[N_{data}(i)-b_mg(i)-tf(i)]cov_{overall}^{-1}(i,j)[N_{data}(j)-b_mg(j)-tf(j)],
\end{align}
\end{linenomath*}
where $cov_{overall}^{-1}(i,j)$ is the inverse of the covariance matrix.
Let $t_0$ be the
result of that fit. The uncertainty is assigned to the $t$ parameter 
by the usual $\Delta \chi^2 = \chi^2_{th(t)}-\chi^2_{\rm best\ fit}$ method.

Now suppose that the background has the form $bf(i)$, where $b$ is a constant
and $f(i)$ is the same known function that appeared for the theoretical model 
parameter.  The
model and background are now completely correlated, that is, they
have the same shape as a function of bin number and differ only
in normalization.  
The minimum $\chi^2$ is then independent of whatever value had been estimated 
for $b_m$, since if $b_m$ is changed, the fitted value
of $t$ will change to compensate, because $f(i)$ is a 
common factor for theoretical model and background. 

Now write the square bracketed terms in the numerator 
for the $\chi^2$ fit, using $b$ not $b_m$, as 
$[N_{data}(i)-(t+b)f(i)]= [N_{data}(i)-zf(i)]$, where $z\equiv t+b$.  
The background experiment has previously obtained
an estimated mean and uncertainty for the parameter $b$.  
The fit here obtains a mean and
uncertainty for the parameter $z=t+b$.

Assume, incorrectly, that the error matrix remains the same as for
the uncorrelated fit. 
If $b$ is estimated by $(b_m)$, the new fit for $t$ is the same
as the old one, $t_{fit}=t_0$.

However, the covariance matrix is not the same as it was for the previous fit.
For this new fit using $z$, the background is part of the
parameter $z$ being fitted. There is no mention of $b$; 
only $z$ appears in the fit. The $z$ fit can be done perfectly well if there
is no prior knowledge of $b$.  The $cov_{B_m}(i,j)$
is not included in the error matrix of the $\chi^2$ terms.
\begin{linenomath*}
\begin{align}
\chi^2 = \sum_{i=1}^{N_{bins}}\frac{[N_{data}(i)-zf(i)]^2}{\sigma^2_{data}}.
\end{align}
\end{linenomath*}

Compare the two results.  If the fit has gone to the same place, both
fits will get $t_0$.  For the $z$ fit, $t=z-b_m$.  
The uncertainty in $t$ will be essentially the
same for both methods.  For the $t$ fit, the uncertainty in the background
appears in the $cov^{-1}_{B_m}(i,j)$ term in the $\chi^2$ formula, 
which broadens the shape of the $\chi^2$
around the minimum value. For the $z$ fit the uncertainty in $t$ due to 
$b$ comes 
explicitly. In $t = z-b_m$, 
the uncertainty in $z$ comes from the $\chi^2$
fit and the uncertainty in $b$ comes from the uncertainty in the measured
value of $b_m$  The difference between the  $t$ fit and the $z$ fit 
comes in the value of $\chi^2$ obtained at
the minimum, the probability of the fit.  For the $t$ fit, even
though the minimum point for $\chi^2$ is independent of the value of $b_m$, the
$\chi^2$ probability of the fit depends on $cov_{B_m}(i,j)$. The $z$ fit gives
a sharper measure of the fit probability.  The $z$ fit asks whether the data
can be fit by an $f(i)$ form, regardless of whether it is background or
signal.  For the $t$ fit, if the background error is sufficiently large,
a good fit can be obtained even when the form of the data differs
very considerably from $f(i)$.

This problem is seen more clearly using the constrained fit method of
Section 2, the ``pulls'' method fit.
\begin{linenomath*}
\begin{align}
\chi^2 = \sum_{i=1}^{N_{bins}}\frac{[N_{data}(i)-(t+b_m+\delta_b)f(i)]^2}
{\sigma^2_{data}(i)}
+\frac{\delta_b^2}{\sigma^2_{b}},
\end{align}
\end{linenomath*}
since $\beta(i)=f(i)$ here.
If there is complete correlation, the pull term will become zero in the fit,
regardless of the size of the actual $\delta_b$.  Any difference between
$b$ and the estimated $b_m$ will be taken up by changing $t$ and leaving
$\delta_b$ zero.  This means that, effectively, one bin has been lost
and the number of degrees of freedom is reduced by one.
The probability obtained using the nominal degrees of freedom 
is then too high, just as seen for the first 
fitting method.

The odd results with both of the $t$, $b$ fits have the same root cause, 
which is seen most easily in the constrained fit method. 
The problem is that $t$ and $b$ are correlated parameters in the fit
appearing only as $t+b$;
$z$ is the only fitting parameter.  The fitting methods
discussed in Section 2 allow correlations between background parameters. 
The constrained fit method
even, as mentioned in Section 2, allows causal correlations between
background and signal parameters such as occur in overall normalization
of the data.  However, it cannot handle a non-causal accidental correlation
between a background parameter and a theory parameter due solely to a
similarity between the shape of the background and theory functions over
the histogram bins, {\it i.e.,}
by a regression correlation.  Fitting $t$ and $b$ without accounting for
this correlation produces a result that is not powerful as the
one obtained when correlations are included.
The fit with $z$ avoids this
problem.  The $z$ fit is a straightforward $\chi^2$ fit with the usual $\chi^2$
properties.  As indicated above, the fit determines the estimate for
$z=t+b$ and the uncertainties of that estimate, while the
preliminary background experiment has obtained the same quantities for $b$.
It is then completely straightforward to combine the two measurements to
get an estimate for the value of $t$ and the uncertainties of that estimate.  
 
The subtlety here is that both $t$ and $b$ are involved in the fit.  
If a fixed value
of $t$ is used and there is no fit, the question of correlation between
$t$ and $b$ does not enter and, for the first method,
the $\chi^2$ distribution should 
include $cov_{B_m}(i,j)$.

The same correlation correction applies for  
the likelihood method. For this discussion, as a shorthand,
$t,\ b,$ and $z$ are taken to be the random 
variables for these parameters.
The variables can be changed from $t$, $b$ to
the independent variables $z$, $b$ with a Jacobian of one.  Then the
probability of the $b$ term can be integrated out and, as with the
$\chi^2$ fits, only $\mathcal{L}(data|z)$ remains
\begin{linenomath*}
\begin{align*}
\mathcal{L}(data|t,b)\mathcal{L}(b) = \mathcal{L}(data|t+b)\mathcal{L}(b)
=\mathcal{L}(data|z)\mathcal{L}(b),\\
\int \mathcal{L}(data|z)\mathcal{L}(b)db =\mathcal{L}(data|z). 
\end{align*}
\end{linenomath*}

Another method to treat the correlation problem would be to make a combined fit for
the background and the model parameters using the combined histograms for 
both the present experiment and
the background experiments.  
However, if there are many backgrounds, often determined
by varied methods, the ``curse of
dimensionality'' makes this impractical.  

Consider the
neutrino oscillation fit, used in MiniBooNE. In this experiment
a beam of muon neutrinos is produced and sent to a detector.  In the
flight from production to interaction in the detector some muon 
neutrinos may have changed into electron neutrinos. The
experiment looks for these electron neutrinos, and tries to determine
the number of electron neutrinos obtained and their energy spectrum. 
The theoretical model for this process (assuming only $\nu_e$ and $\nu_\mu$
neutrinos) involves the product (not sum) 
of two parameters, one a function of $\Delta m^2$ and the other is 
$\sin^22\theta$.
$\Delta m^2$ determines the shape of the energy spectrum and $\sin^22\theta$
is a scale parameter determining the size of the
effect.  A major background comes from muon 
neutrino interactions in the detector
producing $\pi^0$, which decay into two photons and may make the event
appear to be an electron neutrino event.
For the MiniBooNE experiment, does the $\pi^0$ background resemble the 
results expected from the theoretical model? 
$\pi^0$
background numbers for the neutrino exposure are obtained
from Table 6.5 of the MiniBooNE Technical Note 194~\cite{oneninetyfour}.
The neutrino spectrum and the neutrino quasi-elastic cross sections were
read from Figures 2 and 5 from a MiniBooNE neutrino elastic scattering
Physical Review article~\cite{mbelastic}.  
Results are shown in Figure 1.  For
$\Delta m^2= 2$ or $1$ eV$^2$, the mean and $\sigma$ of the  $\pi^0$ background
and the theoretical expectations are quite close and the 
correlations are large. 

\begin{figure}[tbp]
\begin{center}
\vspace{-30mm}
\includegraphics[width =\textwidth]{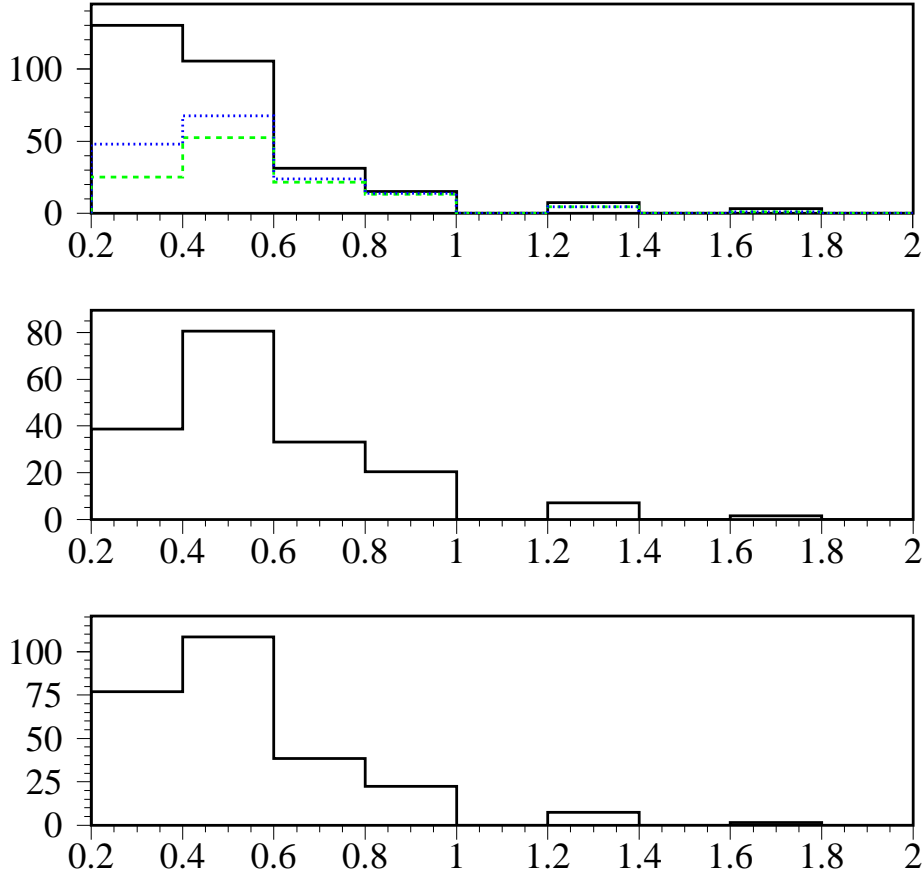}
\vspace{-30mm}
\caption{
The top figure shows the energy spectrum expected
for the  $\pi^0$ background.  The horizontal axis is the energy in GeV obtained
if the $\pi^0$ events are mistakenly reconstructed as charged current
quasi-elastic electron neutrino events.   The dashed green line is the part
of the $\pi^0$ background that is correlated with the theory if
$\Delta m^2 = 0.76$ ev$^2$. The dotted blue line is the part
of the $\pi^0$ background that is correlated with the theory if
$\Delta m^2 = 0.42$ ev$^2$. 
The middle figure shows the energy spectrum
expected for the neutrino oscillation signal if $\Delta m^2 = 0.76$ eV$^2$.
The correlation with the $\pi^0$ spectrum is 0.415.  
The bottom figure shows the energy spectrum
expected for the neutrino oscillation signal if $\Delta m^2 = 0.42$ eV$^2$.
The
correlation with the $\pi^0$ spectrum is 0.587. 
}
\label{Figure 1}
\end{center}
\end{figure}
\section{Partial correlations}
In practice, regression correlations are almost never complete; 
it is necessary to 
consider partial correlations between the theoretical model and background.
In this section a simple model will be considered.  The model then will be
generalized in the following section.
The theory model here will be taken as $N_{th}(i)=tf(i)$, where $f(i)$ is a known
function and $t$ is the fitting variable.  There may be several backgrounds,
but it is assumed here that only background $B_k$ has a significant 
correlation with the theoretical model shape. The background is taken
as $B_k=b_kg_k(i)$, where $g_k(i)$ is a known function and $b_k$ was measured
in a previous experiment with some uncertainty.  Let $M_f$  and $\sigma_f$
be the mean value and spread over the histogram bins for
$f(i)$ and $M_{g_k}$  and $\sigma_{g_k}$ be the mean value 
and spread of $g_k(i)$.
The means and $\sigma$'s here
refer to
mean values and spreads over the bins of the histogram.  They do not
refer to measurement uncertainties.  Let $f'(i)\equiv f(i)-M_f$ and
$g'_k(i)\equiv g_k(i)-M_{g_k}$.  

Start by considering the regression correlation between
$g'_k(i)$ and $f'(i)$.
Consider a plot of the
background $g'_k(i)$ (y-axis) versus the theoretical model 
$f'(i)$ (x-axis), 
where the points are the values for each bin. A straight
line regression
fit of background on theoretical model is made.  Let $x^*_{\rm temp}(i) = af'(i)$, 
 where
$a$  is a  constant chosen to minimize
$\sum_{i=1}^{N_{bins}}[g'_k(i)-x^*_{\rm temp}(i)]^2$. 
The line $x^*_{\rm temp}(i)$ 
represents the completely
correlated part of $g'(i)$ in this linear approximation. The result is
\begin{linenomath*}
\begin{align}
\frac{x^*_{\rm temp}(i)}{\sigma_{g_k}} = \rho\frac{f'(i)}{\sigma_{f}},
\end{align}
\end{linenomath*}
where the regression correlation coefficient is given by
\begin{linenomath*}
\begin{align}
{\rho} = \frac{\frac{1}{N_{bins}}\sum_{i=1}^{N_{bins}}g'_k(i)f'(i)}
{\sigma_{g_k}\sigma_{f}},
\end{align}
\end{linenomath*}
and
%
\begin{linenomath*}
\begin{align}
{\sigma}_{f} = \sqrt{\frac{1}{N_{bins}}\sum_{i=1}^{N_{bins}}(f'(i))^2}.
\end{align}
\end{linenomath*}
$\sigma_{g_k}$ is obtained similarly. This result is derived 
in \cite{probbk} and is discussed in \cite{byron}.

As seen in Equation 9, 
this straight line fit $x^*_{\rm temp}$ is linearly related to the
theoretical model bin dependence $f'(i)$.  For a given theoretical model 
there
is no uncertainty in the values of $M_f$ and $\sigma_{f}$.  With the
present assumption that $g(i)$ is a known function, there is
no uncertainty in the values of $\rho,\ M_{g_k}$ and the value of $\sigma_{g_k}$.
Note that Equations 9 and 10 are
 normalization independent because of
the divisions by $\sigma$.
\begin{linenomath*}
\begin{align} 
x^*_{\rm temp}(i) = \frac{\rho\sigma_{g_k}}{\sigma_f}f'(i) =\rho_{eff}f'(i),
\end{align}
\end{linenomath*}
where $\rho_{eff} \equiv \rho\sigma_{g_k}/\sigma_{f}$ is defined as the
``effective'' correlation coefficient.

$x^*_{\rm temp}(i) $ is appropriate for $f',g'$ not for the full $f,g$ 
which include the mean values.
It does not approach the appropriate limit for full
correlation for $f,g$.  It is necessary to translate along the $f'$ axis.
The size of the translation is uniquely set by asking that, including
mean values, appropriate results are obtained for $\rho=0$ and $\rho=1$.
The proper translation is to plot the points at $f'+\rho_{eff}M_f$. 
The translated $x^*$ is then
\begin{linenomath*}
\begin{align} 
x^*(i)  = x^*_{\rm temp}(i) +\rho_{eff}M_f =\rho_{eff}f'(i) + \rho_{eff}M_f
= \rho_{eff}f(i).
\end{align}
\end{linenomath*}
Since $x^*(i)$ is proportional to $f(i)$, the regression correlation of
$x^*$ with $f$ is one.  It can easily be shown that the regression correlation
of the residuals, $g(i)-x^*(i)$ is zero.
$b_kx^*(i)$ can be considered to be  the correlated 
part of the full background $k$ 
 and the
distances $b_k[g_k(i)-x^*(i)]$ will be taken to  
represent the uncorrelated part
of the background. This division will be assumed to give a reasonable
first order division of the backgrounds.
Let $\Delta_{k}(i)$ be the uncorrelated part of 
background $k$.
\begin{linenomath*}
\begin{align}
\Delta_{k}(i) = b_k[g_k(i)-x^{*}(i)]=b_k\big([g'_k(i) -\rho_{eff}f'(i)] +
[M_g-\rho_{eff}M_f]\big).
\end{align} 
\end{linenomath*}
$\Delta_{k}(i)$ does go to appropriate limits for $\rho=0$ and $\rho=1$.
If $\rho=0$, then $\Delta_{k}(i) = b_kg_k(i)$ as expected.  For $\rho=1$,
then $g'(i)$ is proportional to $f'(i)$, and 
$\rho_{eff} = 1\times (\sigma_{g_k}/\sigma_f)$ is the constant of proportionality.
$ \Delta_{k}(i) = b_k[M_g-(\sigma_{g_k}/\sigma_f)M_f]$.  
For the toy model considered in the previous
section, $g(i)=f(i)$, the means are the same, $\sigma_{g_k}/\sigma_f=1$
and $ \Delta_{k}(i) =0$, as
expected.  If the toy model is extended to be $g(i) = f(i) +C$, where
$C$ is a constant, then the above expression is again correct.

Note that it can turn out that $\rho<0$.  It still will be the proper value
to minimize the sum of the squares of the distances of the background from 
$x^*$.  
Depending on the mean values $M_f$ and $M_g$, the correlated
part of background $k$ 
can have some bins with a value more than the experimental background
points. This should be expected.  Even if the means were the same, 
when the fit to the correlated straight
line is made, some background points will be above and some below that line. 
This is not generally a problem for the method as $b_kx^*(i)+\Delta_{k}(i)$ 
must equal the entire background $B_k(i)$.

The covariance matrix of $\Delta_{k}$ due to the uncertainty in the 
experimental values
obtained for background from the experiments determining the background is
given by:
\begin{linenomath*}
\begin{align}
cov_{\Delta_{k}}(i,j) = [g_k(i)-\rho_{eff}f(i)] 
[g_k(j)-\rho_{eff}f(j)]\sigma^2_{b_k}.
\end{align}
\end{linenomath*}

The fitting function is:
\begin{linenomath*}
\begin{align} 
N_{fit}(i) =tf(i) +b_kx^*(i) 
= zf(i),
\end{align}
\end{linenomath*}
where $z = t +b_k\rho_{eff}$.

$N_{sig}$ in Section 2 had subtracted from data the estimates of all background.
$N_{sig-corr}$ adds to $N_{sig}$ the estimate of the correlated part of $B_k$.
The total background, $B_{tot}(i)=\sum_{j=1}^{N_{bkrd}}[B_j(i)]$, should now
not include all of $B_k(i)$, but should include only the uncorrelated 
part of it.
\begin{linenomath*}
\begin{align}
N_{sig-corr}(i)=N_{data}(i)-\sum_{j\ne k}(B_m)_j(i) -\Delta_{k}(i).
\end{align}
\end{linenomath*}
\begin{linenomath*}
\begin{align}
cov_{corr}(i,j) = cov_{data}(i,j) + cov_{(\Delta_{k}+\sum_{j\ne k}B_j)}(i,j).
\end{align}
\end{linenomath*}
For the first fitting method in Section 2
\begin{linenomath*}
\begin{align}
\chi^2 = 
\sum_{i,j=1}^{N_{bins}}[N_{sig-corr}(i)-zf(i)]
[cov_{corr}^{-1}(i,j)]
[N_{sig-corr}(j)-zf(j)].
\end{align}
\end{linenomath*}
Because the square bracket terms in the $\chi^2$ still add up to data minus 
the sum of theory plus background, 
the only change in the $\chi^2$ calculation due to regression
correlations  
comes in the covariance for this method.

For the constrained fit method, $\beta_{\ell}(i)=\frac{\partial 
[\Delta_{k}(i)+\sum_{j\ne k}(B_m)_j(i)] }
{\partial \delta_{b_\ell}}|_{\delta_{b_\ell}=0}$,
where $\delta_{b_\ell}$ is the deviation
of background parameter $b_\ell$ from its measured value.  Only the
uncorrelated part of the background is used in calculating $\beta_\ell(i)$.
For background $k$, $\beta_k(i) = g_k(i)-\rho_{eff}f(i)$. 

\begin{linenomath*}
\begin{align}
\chi^2 =\sum_{i=1}^{N_{bins}}\frac{[N_{sig-corr}(i) -zf(i)
-\sum_{\ell=1}^{N_{bkrd}}\beta_\ell(i)\delta_\ell]^2}
{\sigma^2_{data}} 
+\sum_{\ell=1}^{N_{bkrd}}
\frac{\delta_{b_\ell}^2}{\sigma^2_{b_\ell}}.
\end{align}
\end{linenomath*}
After the fit the estimate for $t$ is found from
\begin{linenomath*}
\begin{align}
t_{fit} = z_{fit} -\rho_{eff}b_k,
\end{align}
\end{linenomath*}
where, for the first method $(b_m)_k$ is used and for the constrained fit  method
the fitted value of $b_k$ is used.  If the fit was made using $t$ and $b$,
but with the reduced $\Delta_k(i)$, then the value of $t$ is directly obtained
in the fit.
The covariance of $t$ is given by
\begin{linenomath*}
\begin{align}
cov_t = cov_{z_{fit}} + cov_{\rho_{eff}b_k}.
\end{align}
\end{linenomath*}
\section{Generalization of the model}

The correlations considered in the last section can vary with the values
of the theoretical model parameters.  The object of the fit is 
to minimize $\chi^2$
given the correlations relevant to the fit
point.  This means that the correlations must be adjusted to follow the
fitting.  Often the fitting of complex data is done by successive
approximations, such as occur in the popular CERN program Minuit~\cite{minuit}.
Sometimes, the fitting is just
done by calculating $\chi^2$ at points along an $N_t$ dimensional lattice.
However the fit is done, at each step or point, the correlations must be 
recalculated given the model parameters at that step or point.

The assumptions in the last section that the theory could be represented
by $tf(i)$ and the background $k$ by $b_kg_k(i)$ 
with $f(i)$ and $g_k(i)$ 
known functions
were taken to simplify the initial discussion.  Often, theory
and background have more general functions.  Furthermore,
most problems do not have a single correlated 
background parameter and often have more than
one theoretical fitting parameter. It is necessary to include
the effects of multiple backgrounds and multiple theoretical fitting 
parameters. Redo the regression analysis of the last section, 
using the full theoretical model and
the total background instead of $f$ and $g$.  
Let $M_{B_{tot}}$ and $M_{th}$ be the means over the bins of the total
background ($B_{tot}$) and the total theory model ($N_{th}$).  
The $\sigma$'s are defined in
an analogous manner. Let:
\begin{linenomath*}
\begin{align}
B'_{tot}(i) = B_{tot}(i)- M_{B_{tot}};\ N'_{th}(i) = N_{th}(i) - M_{th}.
\end{align}
\end{linenomath*}
Equation 13 becomes:
\begin{linenomath*}
\begin{align}
\frac{x^{*}(i)}{\sigma_{B_{tot}}}= 
\rho\frac{N_{th}(i)}{\sigma_{th}}.
\end{align}
\end{linenomath*}
\begin{linenomath*}
\begin{align}
{\rho} = \frac{\frac{1}{N_{bins}}\sum_{i=1}^{N_{bins}}B'_{tot}(i)N'_{th}(i)}
{\sigma_{B_{tot}}\sigma_{th}}.
\end{align}
\end{linenomath*}

Define $\rho_{eff}\equiv\rho\sigma_{B_{tot}}/\sigma_{th}$. 
\begin{linenomath*}
\begin{align} 
x^*(i) = \rho_{eff}N_{th}(i).
\end{align}
\end{linenomath*}
For independent backgrounds, $x^*(i)$, defined above, is just the sum of
the $x_{k}^*(i)$ for the individual backgrounds $B_k(i)$.  The uncorrelated
part of the background is

\begin{linenomath*}
\begin{align}
\Delta(i) = B_{tot}(i) -x^*(i) = B_{tot}(i)-
 \rho_{eff}N_{th}(i). 
\end{align} 
\end{linenomath*}
\begin{linenomath*}
\begin{align}
N_{sig-corr}(i)=N_{data}(i)-\Delta(i).
\end{align}
\end{linenomath*}
\begin{linenomath*}
\begin{align}
cov_{corr}(i,j) = cov_{data}(i,j) + cov_{\Delta}(i,j).
\end{align}
\end{linenomath*}

If the only uncertainty in the background is the overall normalization $b$ of
the total background, then $cov_{\Delta}(i,j)$ is particularly simple.
For a normalization change, $B_{tot}(i)$, $M_{B_{tot}}$, and 
$\rho_{eff}$ are all proportional to the normalization $b$.  $\rho_{eff}$
is proportional because of the $\sigma_{B_{tot}}$ in the definition.  Therefore
$\delta(\Delta(i)) = (\delta b/b)\Delta(i)$, leading to
$cov_{\Delta}(i,j) = \Delta(i)\Delta(j)\sigma_b^2/b^2$. If there are a sum
of independent backgrounds, each of which has only a 
normalization uncertainty, one can use a simple generalization of this 
calculation since, as was noted above, one can just use the sum
of the individual backgrounds for $x^*$.

Remembering that the bin dependence of $x^*(i)$ is proportional 
to that of 
$N_{th}(i)$ the fitting function can be written
\begin{linenomath*}
\begin{align} 
N_{fit}(i)  = N_{th}(i|\vec{t}\ )
+x^*(i)=N_{th}(i|\vec{t}\ )(1+\rho_{eff}), 
\end{align}
\end{linenomath*}
where $N_{th}(i|\vec{t}\ )=N_{th}(i|t_1,t_2,\cdots,t_{N_{t}})$ 
is the prediction 
for the mean
number of theory model events in bin $i$, given the values of the 
theoretical parameters
$t_1,t_2,\cdots,t_{N_{t}}$. The fitting function should now be viewed as
$N_{fit}$, not just $N_{th}$. In the last section the theory and background
parameters were simple scaling functions.  The regression correlation
considered $f'(i)$ and $g'_k(i)$ which were independent of the scaling.
$\rho_{eff}$ and $M_g$ were not functions of the parameters.  The parameters
$t$ and $b_k$ appeared explicitly in $\Delta$ and $N_{fit}$.  
This simple separation
cannot be assumed for the general problem considered here and the
regression correlation was calculated using the full theory and full
background.  The effect of background parameters and their uncertainties 
now appears in
$N_{fit}$ through $\rho_{eff}$. (If in the last section, the regression
had been done with $tf$ and $bg$ instead of $f$ and $g$, then, because 
$\sigma_{bg}=b\sigma_g$, $\rho_{eff}$ would have been proportional to $b$.)
The $\chi^2$ uncertainty obtained from $\Delta \chi^2$ refers to the
uncertainty in $N_{fit}$.  The uncertainties in $N_{th}$ are obtained
by appropriately convoluting the uncertainties coming from the
$\chi^2$ and the background uncertainties.  This is the generalization
of Equation 22.

For the first fitting method discussed in Section 2,
\begin{linenomath*}
\begin{align}
\chi^2 = 
\sum_{i,j=1}^{N_{bins}}[N_{sig-corr}(i)-N_{th}(i|\vec{t}\ )
(1+\rho_{eff})]\nonumber \\
[cov_{corr}^{-1}(i,j)]
[N_{sig-corr}(j)-N_{th}(j|\vec{t}\ )(1+\rho_{eff})].
\end{align}
\end{linenomath*}
If the constrained fit method is used, assume there are $N_{bkrd}$ background
parameters, $b_k,\ k=1,\cdots,N_{bkrd}$.  Let $\beta_k(i)
=\frac{\partial \Delta(i)}{\partial \delta_k}|_{\delta_k=0}$, 
where $\delta_k$ is
the deviation of background parameter $b_k$ from its measured value.
As in the previous section, only the uncorrelated background is 
used to find $\beta_k(i)$.
\begin{linenomath*}
\begin{align}
\chi^2 = 
\sum_{i=1}^{N_{bins}}\frac{[N_{sig-corr}(i)-N_{th}(i|\vec{t}\ )
(1+\rho_{eff})
-\sum_{k=1}^{N_{bkrd}}\beta_k(i)\delta_k ]^2}
{\sigma^2_{data}}\nonumber\\
+\sum_{k,\ell=1}^{N_{bkrd}}\delta_k[cov^{-1}(k,\ell)]\delta_\ell,
\end{align}
\end{linenomath*}
where $[cov^{-1}(k,\ell)]$ is the inverse of the covariance matrix between
$\delta_k$ and $\delta_\ell$.

\section{Incorporating these correlations in practice}
It was noted
in Section 3, that the fits for neutrino oscillation in the MiniBooNE
experiment are not simple sums.
In terms of the notation from Section 4, for
$tf(i)$, $t$ is determined by  $\sin^22\theta$
and $f(i)$ by a function of $\Delta m^2$.  
The two parameters work together to produce
a single spectrum.  For a background  which matches the shape for a given
 $\Delta m^2$, any mis-estimate of the scale of the background will appear in
the fitted value of $\sin^22\theta$ and the correlated part of the 
background should be associated with that parameter.  

The individual terms in the sum of model terms will sometimes be these
composite terms.  This occurs, for example in fits of neutrino data 
for sterile neutrino 
hypotheses. There will sometimes also be more complicated dependences than the
simple ones here and it is necessary to examine the situation for each
particular experiment as indicated in the last section.  

For determining confidence regions, consider a theoretical model point A. If the
absolute $\chi^2$ is to be used and A
is a fixed point, not the result of a fit, then the regression correlations
should not be included in the calculation of probability.  If
there is a fit, regression correlations should be used. 
The $\Delta \chi^2$ method, 
($\Delta \chi^2 \equiv \chi^2_A - \chi^2_{\rm best\ fit}$) is a 
test often used to determine confidence regions.
The $\chi^2$ at A using the regression
correlations at A minus the best fit  $\chi^2$ using the
regression correlations at the best fit should then be used.
It is important to note that the $\chi^2$ returns the probability
of $N_{fit}$. 

To find a confidence
region for the theoretical parameters, the set of $t$'s, it is, in principle, 
necessary to convolute the $\chi^2$ probability with the probability
for the background uncertainties. 
See the discussion following Equation 30.  If the fit explicitly
included the theory and background, and only modified the covariance
for the correlations, then the theory parameters for the minimum 
are already the appropriate ones, for the first $\chi^2$
method discussed in Section 2.  However, for the constrained fit method
the change in background value must be taken into account.  Calculation
of this change will be discussed in the next section.

For fake data 
studies the experiment is replicated a number of times with Monte Carlo
events.  The theoretical model is held fixed,  However, for each repetition, 
the backgrounds are varied randomly within
their uncertainties. These studies are useful for estimating the expected
variance of the result, for examining the effects of individual backgrounds,
and for examining the sensitivity to the values of the uncertainties of these
backgrounds.  They are widely used, because they do not depend on knowing
the form of psudo-$\chi^2$ probability distributions in advance.

They 
give valid confidence level regions for any valid test.  However, the
use of regression correlations makes the test more powerful as will
be seen in the numerical example in Section 7.
Consider an analogy.  Suppose one estimates the distance between two
objects on a somewhat blurry image, obtaining an estimated distance and 
the uncertainty of that estimate.  If a better focussed image
is available, a new estimate can be made.  Both estimates are correct
given the images each has used, but the estimate from the more
focussed image can be considered a more optimum estimate.

In the presence of the correlations discussed in this note, there is no
change to the usual procedure for choosing Monte Carlo events.  
The regression
correlations do not enter and only the usual backgrounds are randomly varied.

What happens next depends upon the question asked.  Suppose there is no fit
and the question
asked is, ``What is the distribution  of $\chi^2$ if the model parameters
are fixed?''  For this question, the regression correlations do not enter.

However, suppose there is a fit, and it is desired to find the probability
that a specific theoretical model A would give at least as high a
$\chi^2$ as obtained in the real data.  A number of fake data samples are
generated for model A and, for each, the $\chi^2$ for A and the $\chi^2$
values for the best fit are found using the regression correlations.  
Then, if the $\Delta \chi^2=\chi^2(A)-\chi^2({\rm best\ fit})$ method
is used, the likelihood of that model is estimated by the fraction
of times that the Monte Carlo 
$\Delta \chi^2$ is greater or equal to the  $\Delta\chi^2$  
value obtained for the data.
If an absolute $\chi^2$ method is used, then the likelihood is the
fraction of time that the Monte Carlo result is larger than the data $\chi^2$.
If the model and background correlations vary with the model point and
the regression correlations are ignored, the best fit point may be
different. 

At present, the use of an ``effective number of degrees of
freedom'' is frequently employed to give corrections to the nominal number.
The effective number of degrees of freedom is found by fitting the 
$\chi^2$ curve found from fake data studies. 
As noted in Section 3, the presence of systematic
background uncertainties, means that, with or without these corrections,
the probability distribution may not have a $\chi^2$ shape.
With this new procedure the
effective number should be closer to the real number, because the
size of the background uncertainty included in the error matrix is
reduced.
It is
still be useful to include the new effective number as a residual
correction.

\section{An example calculation}

The example taken here is a simplified version of the neutrino oscillation
search in the MiniBooNE experiment.   No MiniBooNE data were used.
$\pi^0$ events are a very major
background to the search for $\nu_e$ events.
 In this example it is taken as the only backround.
The expected background ($\pi^0$-events) 
were taken 
from ~\cite{oneninetyfour}.  
The background uncertainty was taken as a simple scale factor of 15\% 
for the expected number of $\pi^0$ background events in each bin.
The number
of data events was taken to be approximately the same as the number of
events in the MiniBooNE $\nu$ run~\cite{mbelastic}. Eight bins of energy were used.
The energy boundaries for these bins are: 200,300,400,500,
600,800,1000,1500, and 2000 MeV. The true theory parameters were taken
as $\Delta m^2= 0.42$ eV$^2$, 
$\sin^22\theta = 0.01$, for which 
 $\rho=0.587$.  The mean energy, expectation value for 
signal for the data set values and for
background are given in Table 1.  For this example the signal and background
are comparable.

\begin{table}
\begin{center}
\begin{tabular}{||l|l|l|l|l|l|l|l|lr||} \hline
Bin &     1 &     2 &     3 &     4 &     5 &     6 &     7 &     8 & \\ \hline
E & 250.0 &  350.0 &  450.0 &  550.0 &  700.0 &  900.0 & 1250.0 & 1650.0 & \\ 
S &  34.0 &   58.3 &   54.9 &   50.9 &   37.6 &   21.5 &    7.8 &    1.1 & \\
B &  67.8 &   62.4 &   57.7 &   47.8 &   31.2 &   15.3 &    7.5 &    3.5 & \\
\hline \hline
\end{tabular}
\vspace{5mm}
\caption{For each energy bin, the mean energy used E (MeV), the expectation
value for the number of signal (S) events for the data set, 
 $\Delta m^2= 0.42$ eV$^2$, $\sin^22\theta = 0.01$, and the expectation
value for the $\pi^0$ background (B) are given.
}
\end{center}
\end{table}

Confidence regions were calculated using a fake data study 
going over a grid of 21 
points in $\sin^22\theta$ and 41 points in $\Delta m^2$.  
A double loop was made over the grid. 
First each point on the grid was taken in turn as the ``true'' point.  A
set of data was then calculated for that point.  
These events were then analyzed using the expectations for each point on the
grid as the second loop. One thousand twenty four
trials for each point were made. The analysis method used was the constrained
fit method described in Section 2.  The $\delta$ parameter described in
Section 2 was normalized to have an uncertainty with $\sigma=1.$
To do that $\delta$ was multiplied by 0.15 which is the uncertainty
in the background.

In an actual experiment, one uses the measured data set to compare
with the results from the double loop for each of the points on the grid where
the double loop includes statistical and background variations.
In this test the ``data'' was also varied in a data loop 
to find a better average of
results.

A convolution is needed in order to 
determine confidence limits. 
For a test $T$, let $P(T)$ be the probability that a
worse result is obtained for a given value of $T$ (obtained from the
double loop).  Let $f(T)dT$ be the differential probability of $T$ from
the data loop. The quantity required is  $\int P(T)f(T)dT$.  Since the
trials in the data loop follow $f(T)$, this is just the
sum of $P(T_i)/$number of trials. ($T_i$ is the result for trial $i$).

There is a subtle point here.  As noted in Section 4, the fit parameter is 
$z = $ theory plus $\rho_{eff}\times$background.  As noted there,
the $\chi^2$ numerator is still of the form data minus theory minus
background as the sum of the correlated and uncorrelated background
is the full background.  However, when using correlations, the
fit for $\delta$ uses only the uncorrelated part of the background.
This must be transferred to the correlated part of the background.
This shifts the value of $\sin^22\theta$ from the grid value a bit.
$sin^22\theta\rightarrow sin^22\theta(1. - \delta*\rho_{eff}*.15)$.

The constrained fit method includes the
background as another variable to be fit within its known limits.
However, for the fake data discussion
here, Bayes considerations are necessary.  The constrained fit doesn't work
against one background, but works against a different background for each
trial.  Nature has chosen one value for 
the expected background.  If the background is
varied within the double loop, the limits are blurred.  If the background
of the ``data'' sample is also varied, this corresponds to the blurring
occuring twice which tends to hide the difference of the non-correlated
and correlated calculations. This can be seen in Table 2.
The width of the $\delta$ distribution approximately doubles if
background uncertainties are included in the trials.  If the central
background value is one sigma higher than assumed, then the $\delta$ values
are strongly skewed as the constrained fit tries to correct for this.
The one sigma samples are included to illustrate the effect of an
incorrect background assumption, but the analysis will mostly concentrate
on the samples with correct central background estimates.
\begin{table}
\begin{center}
\begin{tabular}{||l|l|l|l|lr||} \hline
$\chi^2$ plot & mean & $\sigma$ & $\delta$ mean & $\delta$ $\sigma$ & \\
\hline \hline
data nc nobkd &  7.25 & 3.69 & -0.0070 & 0.43      & \\
data nc bkd   &  8.06 & 4.18 & 0.0066 & 0.91       & \\
data cor nobkd &  7.87 & 3.86 &  -0.008 & 0.40     & \\
data cor bkd   & 10.83 & 6.54 &  0.007 & 0.83     & \\ \hline
data nc nobkd $1\sigma$ & 8.71 & 4.07 &   0.78 & 0.43 & \\
data nc bkd $1\sigma$ & 9.44 & 4.91 & 0.77 & 0.90 & \\
data cor nobkd $1\sigma$ & 11.239& 5.14 & 0.71 & 0.39 & \\
data cor bkd $1\sigma$ & 14.36 & 9.76 & 0.70 & 0.82 & \\ 
\hline \hline
\end{tabular}
\vspace{5mm}
\caption{Values for the mean and sigma at the data value and
  the mean and sigma of the $\delta$ parameter at the data value.
  Nc means no regression correlations included, cor means regression
  correlations have been
  included. Nobkd means that background expectations were not varied
  in the trials; only statistical variations were used.  Bkd means
  that the trials included both statistical and background variations.
  1$\sigma$ means that the actual background expectation value was taken to
  be one sigma larger than the assumed value.
}
\end{center}
\end{table}

Note that if regression correlations are included the value of the $\chi^2$ is
quite large at the data value if background variations are included.
It was noted earlier that if no fit is
made, but only the $\chi^2$ obtained at a specific value of the theory
parameters, regression
correlations should not be included. If the theory values are
fixed any correlation between background and theory is irrelevant.
If no regression correlations are included, then for no background
variation, the
$\chi^2$ at the data value is lower than the value of eight expected.
Since the background errors are still included in the covariance for the
fit, but not in the Monte Carlo, the results must be low.  They are
less low if regression correlations are included because the size
of the background
variance is decreased.  For full correlation with no background
the $\chi^2$ would be about
one unit low.  Here, $\rho=0.59$ and the difference between the
$\chi^2$ for correlated vs non-correlated fits is 0.62, as was expected.

Table 3 presents the values of the best $\chi^2$ found in going over the
grid of values.  Since the variation of background with trials adds something
not found in Nature, consider the best no background values.
The mean values are close to the expected value of six
for a fit with two parameters, but the widths are a bit smaller than
the expected $\chi^2$ widths of 3.46. The width of the non-corrolated
best $\chi^2$ is close to the value of
3.16 expected for five degrees of freedom.

The mean values are 0.20
larger if regression correlations are included. A back-of-the-envelope
calculation indicates this number is about right.  The error in the
background  is 15\%.  Since the background and theory are approximately
equal this is 7.5\% of the total background plus theory.  For $\rho=0.59$,
the difference between the background error for the non-correlated
and correlated calculations is about 4\%. This would give a difference
of 0.24 of the expected minimum of 6. This difference is
further diluted by the presence of the statistical error in the
error matrix.  The fact that the minima for
both correlated and non-correlated $\chi^2$ is slightly above 6
can be attributed to the fact that the minimum here is a minimum over
the grid points, not the actual minimum.

\begin{table}
\begin{center}
\begin{tabular}{||l|l|l|l|lr||} \hline 
  $\chi^2$ plot & nc mean & nc $\sigma$ & cor mean & cor $\sigma$ & \\
  \hline \hline
best nobkd  & 6.06 & 3.22 &   6.26      & 3.29          & \\
best bkd  &   6.18 & 3.51 &   6.42 & 3.58 & \\ \hline
best nobkd $1\sigma$ & 6.36 & 3.36 & 6.54 &  3.36 &  \\ 
best bkd $1\sigma$ & 6.37 & 3.52 & 6.60 & 3.58 & \\  
\hline \hline
\end{tabular}
\vspace{5mm}
\caption{Values for the mean and sigma at the best value.
    Nc means no regression correlations included, cor means regression
  correlations have been
  included. Nobkd means that background expectations were not varied
  in the trials; only statistical variations were used.  Bkd means
  that the trials included both statistical and background variations.
  1$\sigma$ means that the actual background expectation value was taken to
  be one sigma larger than the assumed value.
}
\end{center}
\end{table}

An important test is to examine the size of the confidence regions. Although
a fake data study will give consistent results with any unbiased
test, a better test will produce a smaller confidence region. If background
variations are included both in the double loop and the data loop, then
the confidence regions are much larger than if they are included in only
one of them.  Here it has been chosen to include the background variations
in the data loop, but not in the double loop.  If the region is chosen
by the $\Delta\chi^2$ test, that is by using $\chi^2 - \chi^2_{best}$, then
if regression correlations are not included the the ratio of sizes of
confidence regions for single variation
to double variation is about 74\% and
about 54\% if regression correlations are included.
If just the simple $\chi^2$ value
is used as a test, then the ratios are 96\% if regression correlations are not
included and 77\% if regression correlations are included.
s noted earlier, the background variation
is really a Bayesian variation.
Putting the variation in the data loop makes a clean
division.  The data loop can be considered
as integrating over the Bayesian prior.  (This method could be followed
with actual data.)  The resulting confidence regions
are not frequentist, but more like Bayesian credible regions.  Results
are shown in Table 4.

The $\Delta$ method gives the smallest confidence regions.  Including
regression correlations significantly lowers the size of the regions.
(A test with only one-fifth the number of
trials was only negligably different from the one used.) 

\begin{table}
\begin{center}
\begin{tabular}{||l|l|l|l|l|l|l|lr||} \hline
Data  &   Method &  nc 90\%  & cor 90\% & ratio & nc 95\% & cor 95\% & 
ratio & \\
\hline \hline
reg. & $\chi^2$  &   54   &     37  &    0.69  &   69   & 54  &  0.78 & \\
reg.  &  $\Delta$ &    27   &     24  &    0.89  & 46  & 37 & 0.80   & \\
\hline
1 $\sigma$ & $\chi^2$  &   43   &    33  &    0.77  &   60  & 40  & 0.67  & \\
1$\sigma$ &  $\Delta$  &   30  &  27   &   0.90 &     36 & 34 &   0.94 & \\
\hline \hline
\end{tabular}
\vspace{5mm}
\caption{Table of the number of grid points within the confidence limits.
1$\sigma$ means the background is higher than estimated by $1\sigma$ (15\%).
$\chi^2$ means that
the test is just the $\chi^2$ value for the data.  
$\Delta$ means that the test is $\chi^2-\chi^2_{best}$. Nc means that 
regression correlations were not included and cor means that they were included.
Ratio is the ratio of number of grid points  in the confidence region
included for cor divided by
the number included for nc.  The double
loop included only statistical variations.
The ``data'' consisted of 5001
repetitions with both statistical and background variations.  
} 
\end{center}
\end{table}

The $\chi^2$ are shown for the data value
with and without regression correlations in Figure 2.
The $\chi^2$ plot with regression correlations has a larger tail.
Figure 3 shows the $\chi^2$ plot with and without regression correlations
for the point on the  grid corresponding to the best $\chi^2$ value.
The means
and $\sigma$'s for these plots were given in Tables 2 and 3.

Comparison of the confidence
regions obtained with and without including regression correlations are
shown in Figures 4-7.
The 90\% and 95\% confidence level plots are
shown in Figure 4 for the $\chi^2$ test and Figure 5 for the $\Delta$ test.  
Results are shown in Figures 6 and 7 for the sample with the background 15\% 
higher than the assumed background.

\begin{figure}[tbp] 
\begin{center}
\vspace{-30mm}
\includegraphics[width =\textwidth]{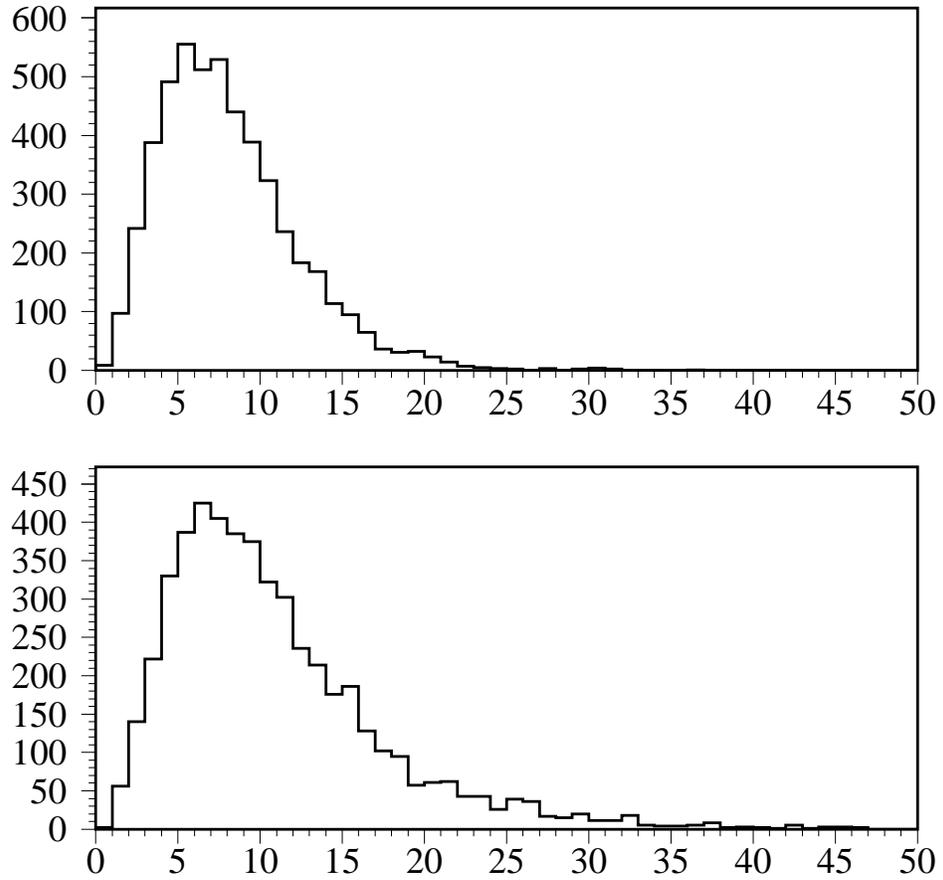}
\vspace{-30mm}
\caption{    
The $\chi^2$ distribution is shown for the data values.
The top
plot shows the distribution if the regression correlation method is not
used and the bottom plot shows the result if the regression correlations
are included. Both statistical and background variations are included.
}
\end{center}
\label{Figure 2}
\end{figure}

\begin{figure}[tbp] 
\begin{center}
\vspace{-30mm}
\includegraphics[width =\textwidth]{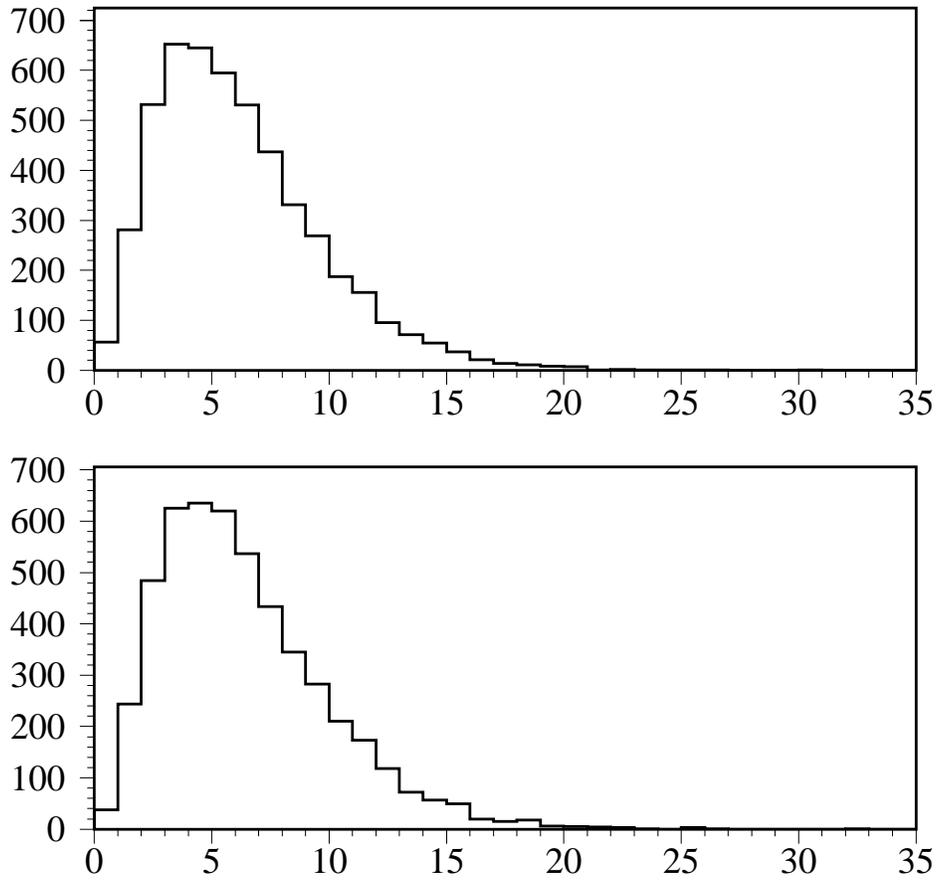}
\vspace{-30mm}
\caption{
The best $\chi^2$ obtained in going over the grid of
points is shown if the data was fit using the data values.
The top
plot shows the distribution if the regression correlation method is not
used and the bottom plot shows the result if the regression correlations
are included. Both statistical and background variations are included.
}
\end{center}
\label{Figure 3}
\end{figure}

\begin{figure}[tbp] 
\begin{center}
\vspace{-30mm}
\includegraphics[width =\textwidth]{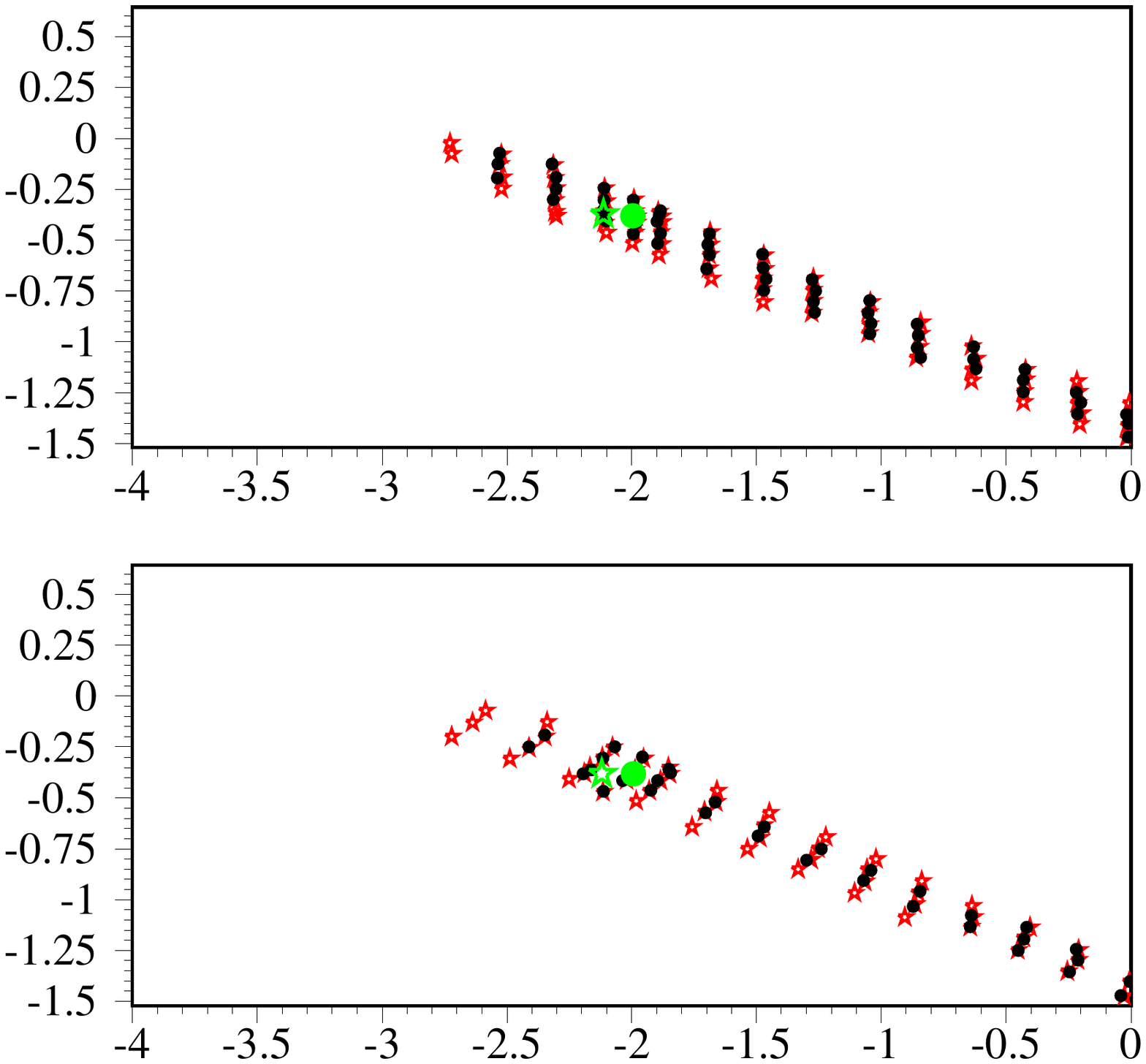}
\vspace{-30mm}
\caption{
Plots of $\Delta m^2$ vs $\sin^22\theta$ for events within
90\% and 95\% confidence regions using the ``$\chi^2$'' method.  The
x-axis is the log$_{\rm 10}$ of $\sin^22\theta$ and the y-axis is the
log$_{\rm 10}$ of  $\Delta m^2$. The top
plot shows the acceptance if the regression correlation method is not
used and the bottom plot shows the result if the regression correlations
are included.  The black dots are within the 90\% confidence region and
the red stars are within the 95\% region.  The open green star is the best fit
point and the filled green circle  is the ``real'' data point.
}
\end{center}
\label{Figure 4}
\end{figure}
 
\begin{figure}[tbp] 
\begin{center}
\vspace{-30mm}
\includegraphics[width =\textwidth]{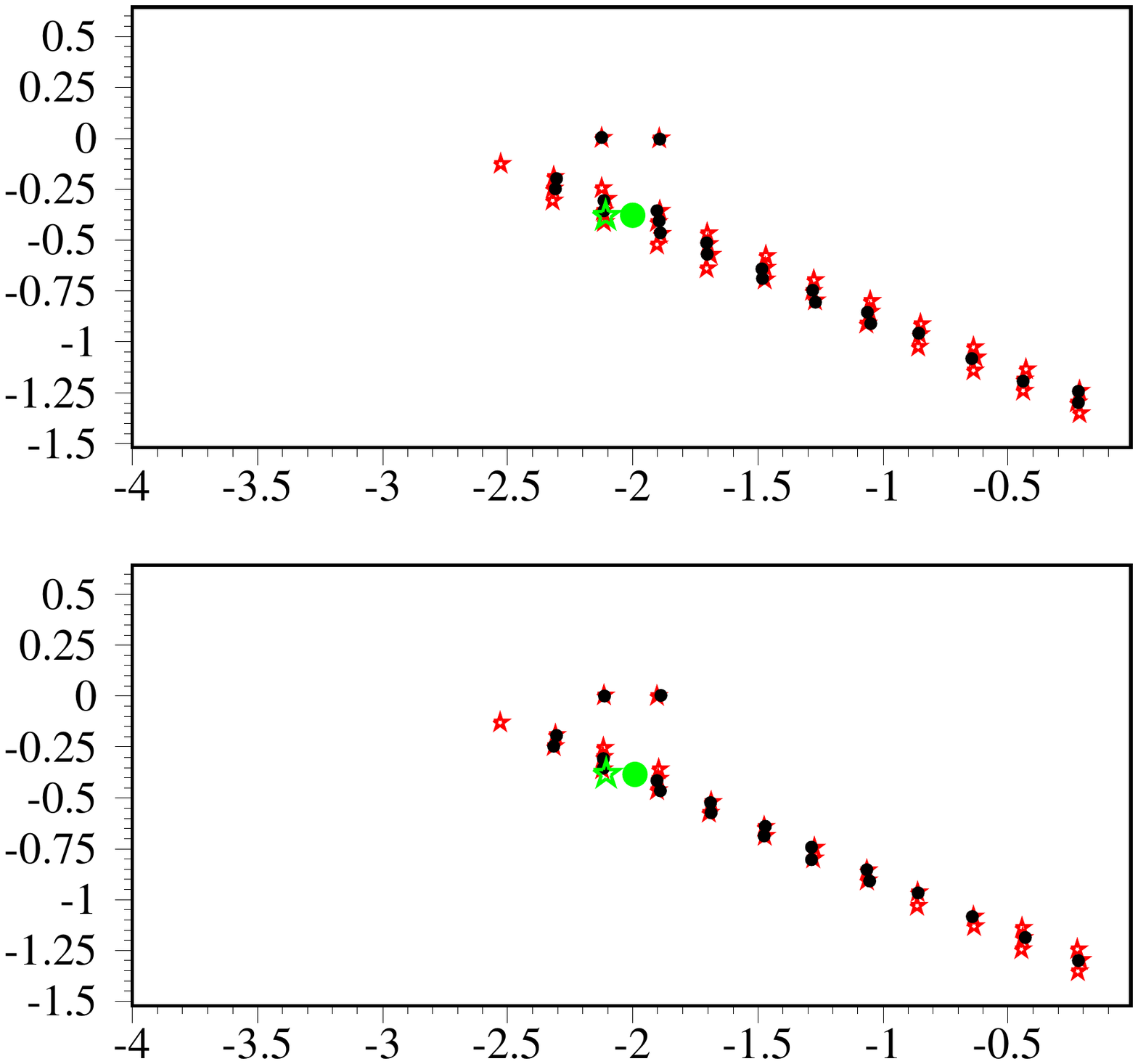}
\caption{
Plots of $\Delta m^2$ vs $\sin^22\theta$ for events within
90\% and 95\% confidence regions using the ``$\Delta$'' method.  The
x-axis is the log$_{\rm 10}$ of $\sin^22\theta$ and the y-axis is the
log$_{\rm 10}$ of  $\Delta m^2$. The top
plot shows the acceptance if the regression correlation method is not
used and the bottom plot shows the result if the regression correlations
are included.  The black dots are within the 90\% confidence region and
the red stars are within the 95\% region.  The open green star is the best fit
point and the filled green circle  is the ``real'' data point.
}
\end{center}
\label{Figure 5}
\end{figure}

\begin{figure}[tbp] 
\begin{center}
\vspace{-30mm}
\includegraphics[width =\textwidth]{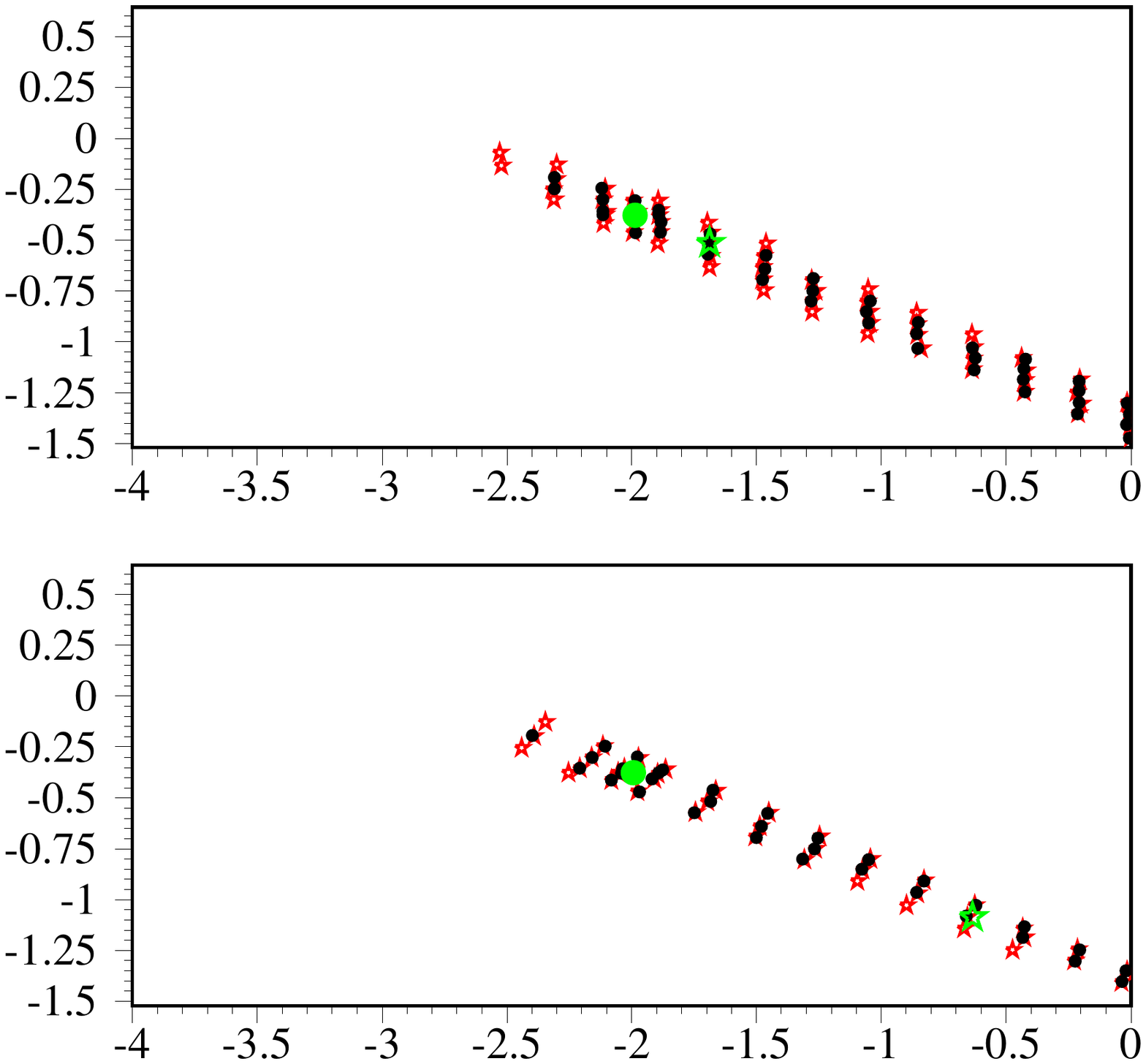}
\caption{
Plots of $\Delta m^2$ vs $\sin^22\theta$ for events within
90\% and 95\% confidence regions using the ``$\chi^2$'' method.  It is 
assumed for these plots that the background is 15\% larger than expected. The
x-axis is the log$_{\rm 10}$ of $\sin^22\theta$ and the y-axis is the
log$_{\rm 10}$ of  $\Delta m^2$. The top
plot shows the acceptance if the regression correlation method is not
used and the bottom plot shows the result if the regression correlations
are included.  The black dots are within the 90\% confidence region and
the red stars are within the 95\% region.  The open green star is the best fit
point and the filled green circle  is the ``real'' data point.
}
\end{center}
\label{Figure 6}
\end{figure}

\begin{figure}[tbp] 
\begin{center}
\vspace{-30mm}
\includegraphics[width =\textwidth]{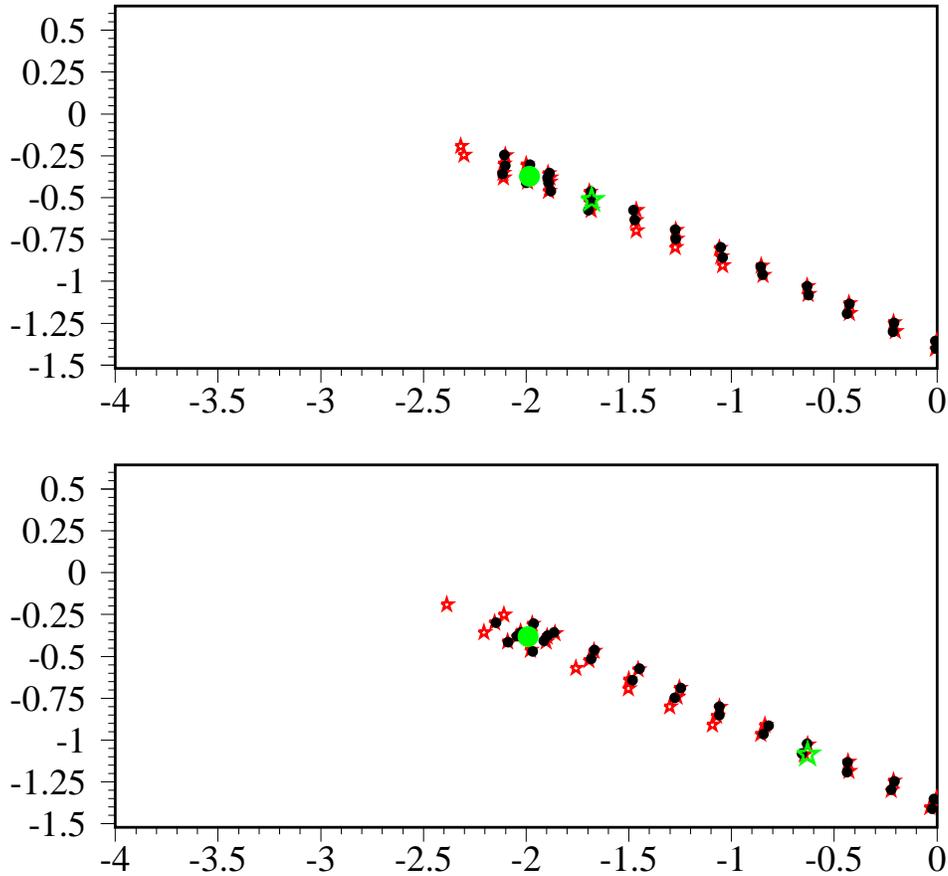}
\caption{
Plots of $\Delta m^2$ vs $\sin^22\theta$ for events within
90\% and 95\% confidence regions using the ``$\Delta$'' method.  It is 
assumed for these plots that the background is one $\sigma$ (15\%) larger 
than expected. The
x-axis is the log$_{\rm 10}$ of $\sin^22\theta$ and the y-axis is the
log$_{\rm 10}$ of  $\Delta m^2$. The top
plot shows the acceptance if the regression correlation method is not
used and the bottom plot shows the result if the regression correlations
are included.  The black dots are within the 90\% confidence region and
the red stars are within the 95\% region.  The open green star is the best fit
point and the filled green circle  is the ``real'' data point.
}
\end{center}
\label{Figure 7}
\end{figure}
\clearpage

\section{Summary}
Methods are given for using the $\chi^2$ method when regression 
correlations between the shapes of
backgrounds and theoretical models being fitted occur.
These methods  
are appropriate whenever these correlations exist.
 
Fake data studies without including these correlations often will not
be optimum.  The use of ``effective number of degrees of freedom'' 
will help the
situation, but will not be as accurate as the methodology introduced here.
If regression correlations are used, the confidence regions are smaller than
if they are not used.



\section{Acknowledgements}
I wish to acknowledge the considerable help of Xuming-He, the H. C. Carver
Professor of Statistics, University of Michigan, during several
discussions which helped clarify
the problem and correct several errors.  Michael Shaevitz, Professor of
Physics, Columbia University, Janet Conrad, Professor of Physics, MIT, 
MIT graduate student 
Christina Ignarra and
LANL Staff Member William Louis read through the note  and helped
to find ambiguities and errors. 


\clearpage

{}

\end{document}